\documentclass[%
 reprint,
superscriptaddress,
nofootinbib,
 amsmath,amssymb,
 aps,
 prd,
]{revtex4-2}

\usepackage{graphicx}% Include figure files
\usepackage{dcolumn}% Align table columns on decimal point
\usepackage{bm}% bold math
\usepackage{xcolor}
\usepackage{orcidlink}

\DeclareRobustCommand{\VAN}[3]{#2}
\let\VANthebibliography\thebibliography
\def\thebibliography{\DeclareRobustCommand{\VAN}[3]{##3}\VANthebibliography}

\newcommand{\msun}{{\,\rm M_\odot}}

\newcommand{\kms}{\,{\rm km}\,{\rm s}^{-1}}

\newcommand{\Gyr}{\,{\rm Gyr}}

\newcommand{\cpm}{\,{\rm cm}^2\,{\rm g}^{-1}}

\def\jcap{J. Cosmol.  Astropart. Phys.}

\def\apj{ApJ}

\def\apjl{ApJ}
\def\mnras{MNRAS}
\def\araa{ARA\&A}
\def\aj{AJ}

\def\physrep{Phys. Rep.}
\def\nat{Nature}

\def\apjs{ApJS}
\def\prd{Phys. Rev. D}

\begin{document}

\preprint{APS/123-QED}

\title{Core collapse in resonant self-interacting dark matter across two decades in halo mass}% Force line breaks with \\
%\thanks{}

\author{Vinh Tran\,\orcidlink{0009-0003-6068-6921}}
 \affiliation{Department of Physics and Kavli Institute for Astrophysics and Space Research, Massachusetts Institute of Technology, Cambridge, MA 02139, USA}
 \email{vinhtran@mit.edu}

\author{Xuejian Shen\,\orcidlink{0000-0002-6196-823X}}
\affiliation{Department of Physics and Kavli Institute for Astrophysics and Space Research, Massachusetts Institute of Technology, Cambridge, MA 02139, USA}
 
\author{Daniel Gilman\,\orcidlink{0000-0002-5116-7287}}
\affiliation{Department of Astronomy \& Astrophysics, University of Chicago, Chicago, IL 60637, USA}
\affiliation{Brinson Prize Fellow}

\author{Mark Vogelsberger\,\orcidlink{0000-0001-8593-7692}}
\affiliation{Department of Physics and Kavli Institute for Astrophysics and Space Research, Massachusetts Institute of Technology, Cambridge, MA 02139, USA}

\author{Stephanie O'Neil\,\orcidlink{0000-0002-7968-2088}}
\affiliation{Department of Physics \& Astronomy, University of Pennsylvania, Philadelphia, PA 19104, USA}
\affiliation{Department of Physics, Princeton University, Princeton, NJ 08544, USA}

\author{Donghua Xiong}
\affiliation{Tandon School of Engineering, New York University, New York, NY 11201, USA}

\author{Jiayi Hu}
\affiliation{Department of Mathematics, Columbia University, New York, NY 10027, USA}

\author{Ziang Wu}
% \affiliation{Department of Mathematics, Tandon School of Engineering, New York University}
\affiliation{Tandon School of Engineering, New York University, New York, NY 11201, USA}

\date{\today}% It is always \today, today,

\begin{abstract}
    Core collapse, a process associated with self-interacting dark matter (SIDM) models, can increase the central density of halos by orders of magnitude with observable consequences for dwarf galaxy properties and gravitational lensing. Resonances in the self-interaction cross section, features of hidden-sector models with light mediators and attractive potentials, can boost the strength of self-interactions near specific relative velocities, accelerating collapse in halos with central velocity dispersions near the resonance. To explore this phenomenon, we present a suite of idealized N-body simulations of isolated halos with masses $10^7 \text{--} 10^9\msun$ evolved under two resonant cross section (RCS) models with localized enhancements to the cross section on scales $v \sim 5 \text{--} 50 \ \rm{km} \ \rm{s^{-1}}$. We show that the change in halo internal structure depends on how the velocity distribution of bound particles moves across resonances in the cross section during core formation and collapse. The interplay between the velocity distribution of bound particles and localized features of the cross section causes deviations from self-similar evolution, a characteristic of velocity-independent cross sections, at the level of up to $20\%$. Depending on the alignment with resonant features, halos of different masses reach different evolutionary stages after a fixed physical time and develop diverse density profiles and rotation curves.
    %We show that the self-similar evolution observed in models with velocity-independent cross sections approximately holds for RCS models during core formation, provided the resonance {\bf{DG: I am trying to make this a little less technical}} the heat conductivity-averaged effective cross section $\sigma_\kappa$ changes gradually with respect to the core one-dimensional velocity dispersion $\sigma_{\rm{1D,c}}$, i.e. $\lvert \partial \ln{\sigma_\kappa} / \partial \ln{\sigma_{\rm{1D,c}}} \rvert \lesssim 1.5$. More prominent resonances ($\gtrsim 2$) cause self-similarity to break down at the level of $\sim 10\%$ in the core density and radius. In both RCS models, the core-collapse times of halos deviate from the universal track by $\sim 5\text{--}15\%$, varying on a halo-by-halo basis. Depending on the alignment to resonant features, halos evolved under different effective cross sections reach different evolution stages, developing denser or shallower profiles. These processes could contribute to the diversity of galactic rotation curves predicted by SIDM models.
    
\end{abstract}

%Use showkeys class option if keyword
\maketitle

%\tableofcontents

\section{Introduction}
\label{sec:intro}

The standard cosmological paradigm based on collisionless cold dark matter (CDM) has proven quite successful in explaining the large-scale structures in the Universe and provides the foundation to understanding galaxy formation~\citep{Blumenthal1984,Davis1985}. However, it encounters challenges at small astrophysical scales~\citep[see e.g. the review ][]{Bullock2017}, in particular in explaining the observational properties of dwarf galaxies. For instance, while observed dwarf spheroidal galaxies and low-surface brightness galaxies commonly exhibit cored central density profiles~\citep[e.g.,][]{Flores1994,Moore1994,deBlok2001,KDN2006,Oh2011,Oh2015}, CDM-only simulations typically produce a universal cuspy profile at halo center~\citep[e.g.][]{Navarro1996,Navarro1997,Moore1999,Klypin2001}. On the other hand, challenges arise from the population of massive, concentrated subhalos found in simulations, which are inconsistent with the stellar kinematics of observed satellite galaxies around the Milky Way or M31~\citep{MBK2011,MBK2012,Tollerud2014}. These discrepancies have recently coalesced into the so-called ``diversity problem'', where the rotation curves of dwarf galaxies -- both in the field \citep{Oman2015} and among the Milky Way’s satellites \citep{Kaplinghat2019} -- display greater diversity than CDM prediction.

Consequently, it is crucial to investigate non-standard DM models that may resolve these small-scale tensions. One possibility is self-interacting dark matter (SIDM), a class of models proposed and studied over the past three decades \citep[e.g.,][]{Carlson1992, deLaix1995, Firmani2000,Spergel2000}. These models are often motivated by hidden dark sector extensions to the Standard Model~\citep[e.g.,][]{Ackerman2009,Arkani-Hamed2009,Feng2009,Feng2010,Loeb2011,vandenAarssen2012,CyrRacine2013,Tulin2013,Cline2014}. SIDM can potentially address small-scale issues \citep[see the review by][and references therein]{Tulin2018} by enabling efficient heat conduction, which yields isothermal, cored density profiles at halo centers \citep[e.g.,][]{Vogelsberger2012, Rocha2013, Zavala2013, Elbert2015,Shen2022-sidm}. SIDM halos also show stronger responses to the variations in both halo concentrations and the gravitational potential provided by baryonic matter, offering a possible explanation for the diversity problem \citep[e.g.,][]{Kamada2017,Creasey2017,Omid2020,Oman2015,Kaplinghat2019,Jiang2023}. 

One key prediction of SIDM models is that some fraction of halos should experience the ``gravothermal catastrophe''~\citep[e.g.,][]{Bell1968,Bell1980} and ultimately collapse into structures with extremely high central densities \citep[e.g.,][]{Burkert2000,Kochanek2000,Balberg2002,Colin2002,Koda2011,Vogelsberger2012,Elbert2015,Correa2021,Shen2021,Slone+23}. These core collapsed halos could contribute to the diversity of galactic rotation curves, and have pronounced gravitational lensing signatures \citep[e.g.][]{Gilman+21}. 

Most studies of SIDM models in the context of core collapse have focused on either constant cross sections or relatively simple, monotonic velocity dependencies. In recent years, the scope of investigation has broadened to encompass a variety of cross sections that emerge from more complex dark sectors~\cite{Tulin_2013,Tulin13,Chu++20,Gilman+23,Kamada+24}. In \cite{Tran2024_1} and this follow-up paper, we examine a family of SIDM models featuring resonances in the cross section, leading to an order-of-magnitude enhancement of scattering near the characteristic relative velocity. These features can manifest as pronounced peaks, multiple peaks, or broader regions of suppression or amplification across different velocities. In these scenarios, the cross section amplitude can vary substantially throughout a halo’s radius, with the implication that the cross section strength ``seen'' by a halo changes as a function of radius when pronounced resonances are present. 

In~\citep{Tran2024_1}, we investigated the gravothermal collapse of SIDM halos exhibiting a single strong resonance feature in the cross section by performing idealized N-body simulations of isolated halos. In this paper, we extend the analysis to halos with multiple resonances in the cross section, and cover a wider range of halo masses. We have two main goals: First, we will assess more comprehensively how the evolution of a halo with a resonant cross section depends on the central velocity dispersion of the halo, as determined by the halo mass. Second, we will examine to what degree self-similar structural evolution seen in SIDM models with velocity-independent cross sections is also seen in those exhibiting single and multiple resonant peaks. 

This paper is organized as follows: Section \ref{sec:particle_physics} introduces the particle physics mechanisms behind resonances in the cross section. Section \ref{sec:simulation} details our simulation setup and the distribution of particle collision velocities inside halos. Section \ref{sec:results} presents the results of our simulations, focusing on the evolutions of halo core density, core velocity dispersion, core half-density radius, as well as the rotation curves of halos. Section \ref{sec:conclusions} summarizes our findings and discusses the implications of the results of resonance features.

\section{Particle physics model}
\label{sec:particle_physics}

% \begin{figure}
%     \centering
%         \includegraphics[width=0.49 \textwidth]{Figures/resonant_cross_section_models.png}
%     \caption{The cross section profiles of SIDM models with single (blue) and multiple (pink) resonant peaks. Both models are normalized to have an amplitude of $\sigma_V/m = 1\cpm$ at the relative velocity $v_{\rm{rel}} = 200\kms$. The vertical dashed lines show the peak of the halo core relative velocity distributions $p^{\rm{core}}_{v_{\rm{rel}}}$ (following Equation \ref{eqn:core_relative_velocity_df}, and shown in Figure \ref{fig:relative_velocity_structure}) for halos of mass $10^{6.5}$ to $10^{10}\msun$.}
%     \label{fig:resonant_cross_section}
% \end{figure}

% \begin{figure}
%     \includegraphics[width=0.49 \textwidth]{Figures/single_peak_ell.pdf}
%     \includegraphics[width=0.49 \textwidth]{Figures/multi_peak_ell.pdf}
%     \caption{The single and multi-peak cross sections calculated as functions of $\ell_{\rm{max}}$ (see Equation \ref{eqn:viscositycross}). The single-peak model arises from the $\ell = 1$ term in the summation over partial waves, while the multi-peak model has higher-order resonances with contributions up to the $\ell = 6$. }
%     \label{fig:elldep}
% \end{figure}

We consider self-interactions between DM particles of mass $m_{\chi}$ through a light mediator of mass $m_{\phi}$. As in Paper\,I~\citep{Tran2024_1}, we consider an attractive Yukawa potential
\begin{equation}
\label{eqn:yukawapot}
V\left(r\right) = - \dfrac{\alpha}{r}\exp{\left(-r\, m_{\phi}\right)}
\end{equation}
where $\alpha$ sets the strength of the interaction, and $m_{\phi}$ is the mass of a light mediator particle. We compute the cross section for this potential using partial wave analysis. The differential scattering cross section is given by
\begin{equation} 
\label{eqn:partialwavesum}
\frac{{\rm d} \sigma}{{\rm d} \Omega} = \frac{1}{k^2} \Big | \sum_{\ell = 0}^{\infty} \left(2 \ell + 1\right)e^{i \delta_l} \sin \delta_l \,P_{\ell}\left(\cos\theta\right)\Big |^2
\end{equation}
where the phase shifts $\delta_{\ell}$ encode the evolution of the wave function in the potential $V\left(r\right)$. $k = m_{\chi} v /2$ and $v$ represent the magnitudes of the momentum and velocity. To determine the phase shifts, we solve the differential equation \citep{Chu++20}
\begin{multline}
\dfrac{\partial \delta_{\ell}\left(r\right)}{\partial r} = -k\, m_{\chi}\, r^2\, V\left(r\right)\, \big[ \cos\left(\delta_{\ell}\left(r\right)\right)\, j_{\ell}\left(kr\right) \\
- \sin\left(\delta_{\ell}\left(r\right)\right)\, n_{\ell}\left(kr\right)\big]^2,
\end{multline}
with the boundary condition $\delta_{\ell}\left(0\right) = 0$, and obtain $\delta_{\ell}$ from taking the limit $r \rightarrow \infty$. 

In our numerical implementation of the scattering process, we follow common practice and use a proxy for the differential scattering cross section ${\rm{d}} \sigma / {\rm{d}} \Omega$ that averages over the angular dependence \citep{Yang+2022}
\begin{eqnarray}
\label{eqn:viscositycross}
\sigma_V &=& \int \frac{{\rm d} \sigma}{{\rm d} \Omega} \sin^2\theta {\rm d} \Omega.
\end{eqnarray}
The expression for $\sigma_V$ in terms of the phase shifts is given by \citep{Colquhoun+21}
\begin{equation}
\label{eqn:viscositycross}
\sigma_V = \frac{4 \pi}{k^2} \sum_{\ell = 0}^{\ell_{\rm{max}}} \frac{\left(\ell + 1\right)\left(\ell +2\right)}{2 \ell + 3} \sin^2\left(\delta_{\ell +2} - \delta_{\ell}\right).
\end{equation}

The two cross sections we consider are shown in Figure \ref{fig:cross_sections}. The first model, which from now on we refer to as the \textit{single-peak model}, is shown in the upper panel. This is the same cross section considered in Paper\,I, with $m_{\rm{\chi}} = 31.8 \ \rm{GeV}$, $m_{\rm{\phi}} = 5.7 \ \rm{MeV}$, and $\alpha = 0.00158$. The second model, has $m_{\chi} = 67.7 \ \rm{GeV}$ $m_{\phi} = 1.9 \ \rm{MeV}$, and $\alpha = 0.00155$; we refer to this as the \textit{multi-peak model}. For the single-peak model, the $\ell = 1$ partial wave produces the resonant structure and higher order partial waves do not contribute, while the structure of the multi-peak cross section depends on partial waves contributions up to $\ell = 6$. The single and multi-peak cross sections are two examples of a rich variety of SIDM models that can affect the evolution of small-scale cosmic structure \cite[e.g.][]{Tulin_2013,Gilman+23}. We have chosen these two examples to understand how the evolution of halo structure depends on the width of a resonance in the cross section.

These cross sections exceed $100 \ \rm{cm^2} \rm{g^{-1}}$ at relative velocities below $20 \ \rm{km} \ \rm{s^{-1}}$, high enough to trigger core-collapse in low-mass halos \citep[e.g.][]{Correa2021,Sameie2020,Nadler+23}. The cross section is $\sigma_V / m_{\rm{\chi}} = 1 \cpm$ at galactic scales $v\sim200\kms$, and drops as $v^{-4}$ towards higher relative velocities, evading upper limits from galaxy clusters at $v \sim 1000 \ \rm{km }\ \rm{s^{-1}}$ \cite[e.g.][]{Peter+2013}. We note that some SIDM models with light mediators, such as those considered in this work, can also produce distinct features on cosmological scales through a suppression of the linear matter power spectrum \cite[e.g.][]{Vogelsberger16,Nadler++24}. In some scenarios, this is an additional feature of SIDM that should be considered when running full cosmological simulations. Whether or not a given self-interaction cross section is accompanied by a suppression of small-scale power is model-dependent. In this work, we assume a particular model for the elastic scattering cross section and use this model to compute the rate of heat transfer in halos. Other considerations, such as inelastic processes that possibly occur alongside elastic scattering, and any possible suppression of the matter power spectrum, depend on specific details of the particle physics model. Our assumed inelastic scattering cross sections correspond to a dark sector model for which there is negligible suppression of the matter power spectrum on the scales of interest.

\section{Simulations}
\label{sec:simulation}

In this section, we discuss the setup of our idealized N-body simulations and outline the analysis process. We begin in Section \ref{subsec:distributions} by examining the distribution of particle collision velocities inside halos with the RCS SIDM models discussed above and introducing velocity-averaged self-interaction cross sections that determine the halo evolution. These analyses serve as the basis for our choice of halo masses of simulations. In Section \ref{subsec:numerical_setup}, we review the technical aspects of the simulations and present the setup of our halos. Section \ref{subsec:analysis} details our approach in analyzing the simulation outputs.

\subsection{Collisional distribution functions}
\label{subsec:distributions}

\begin{figure}
    \centering
    \includegraphics[width=0.49 \textwidth]{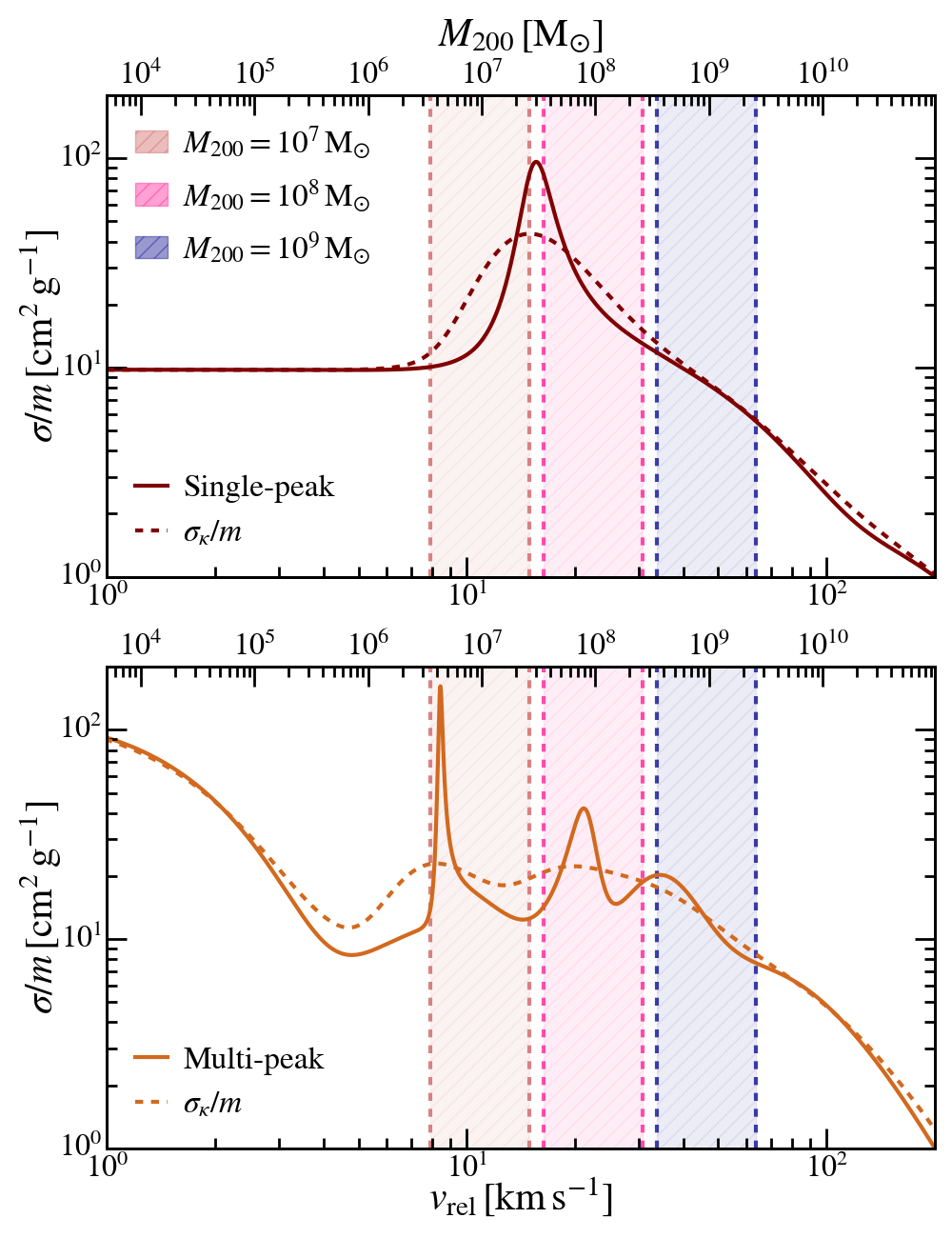}
    \caption{The single- (top) and multi-peak (bottom) cross sections per unit of mass ($\sigma/m$, solid lines) versus the relative velocity of DM particles and halo mass, which are matched based on the most probable relative velocity magnitude $v_{\rm{rel},\kappa}^{\rm{max}}$ of the heat conductivity-weighted MB distribution for each halo (Equation \ref{eqn:p_kappa}). The dashed lines show the heat conductivity-averaged cross sections $\sigma_\kappa/m$, and the shaded regions show the FWHM of the heat conductivity-weighted MB distributions for the halos of mass $10^7, 10^8$, and $10^9\msun$. Although the resonant features of the multi-peak RCS model are more complicated, in the relevant scales, $\sigma_\kappa/m$ appears flatter than that in the single-peak model.}
    \label{fig:cross_sections}
\end{figure}

\begin{figure}
    \centering
    \includegraphics[width=0.49 \textwidth]{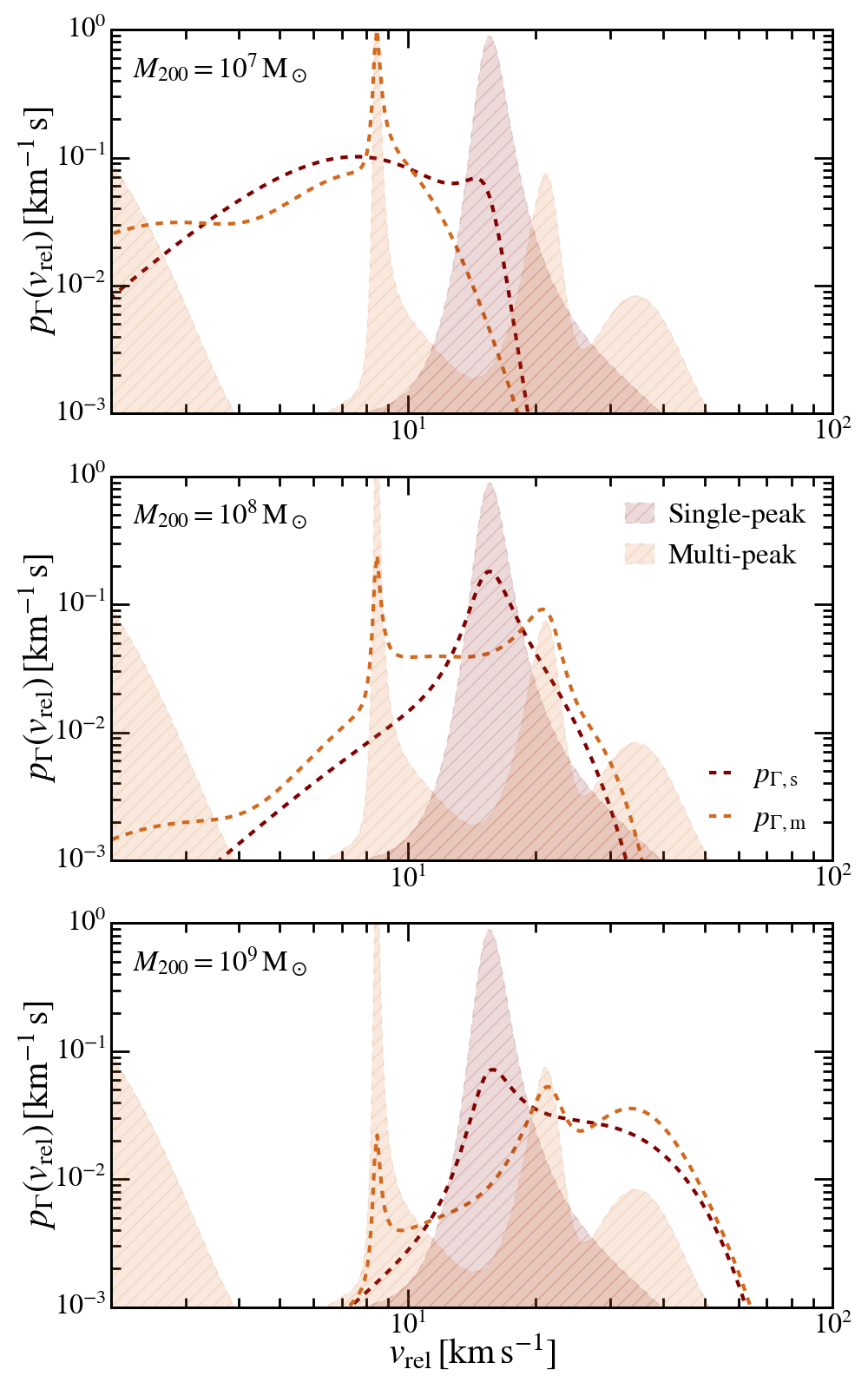}
    \caption{The collision rate-weighted MB distributions calculated using the single-peak (red) and multi-peak (orange) RCSs, $p_{\Gamma,{\rm{s}}} (v_{\rm{rel}})$ and $p_{\Gamma,{\rm{m}}} (v_{\rm{rel}})$, respectively, following Equation \ref{eqn:p_Gamma} with $\sigma_{\rm{1D}} = 1.10 \sigma_{\rm{s}}$ for the halos of mass $10^7$ (top), $10^8$ (middle), and $10^9\msun$ (bottom). The cross section profiles (shaded fills) are also shown as references. As the halo mass changes, the principle resonant features governing the particle collisions shift. These features deviate significantly from the regions most relevant to the heat conductivity-weighted MB distributions and corresponding timescale, which is as expected from the constructions of the two distributions.}
    \label{fig:relative_velocity_distribution}
\end{figure}

The characteristic core-collapse timescale of SIDM halos with elastic self-interactions is~\citep{Yang2024}
\begin{equation}
    \label{eqn:collapse_timescale}
    \tau(\sigma_{\rm{eff}}/m) = \frac{150}{C} \frac{1}{\sigma_{\rm eff}/m} \frac{1}{\rho_{\rm{s}}} \left(\frac{1}{4 \pi \, \sigma_{\rm{s}}^2}\right)^{1/2},
\end{equation}
where $\sigma_{\rm{s}} \equiv \sqrt{G \, \rho_{\rm{s}} \, r_{\rm{s}}^2}$. $\rho_{\rm{s}}$ and $r_{\rm{s}}$ are some choice of characteristic scale density and scale radius, which are often chosen to be those of the NFW profile~\citep{NFW1997}. Here, $C$ is an order-unity prefactor that is chosen by convention to match the evolution of halos in N-body simulations and analytic gravothermal fluid models \cite{Essig2019}. We use $C \simeq 0.85$\footnote{In Paper\,I~\citep{Tran2024_1}, we mistakenly reported all values of $C$ as 1.5 times smaller than the actual values. However, this is of no consequence to our analysis, as the relative comparisons between models are not affected by the absolute values.} following our previous work~\citep{Tran2024_2}. We define the scaled time as $t = T/\tau(\sigma_{\rm eff}/m)$, where $T$ is the physical time. As $\sigma_{\rm eff}/m$ can vary when the halo evolves, we compute $t$, instead, by integrating the instantaneous ``phase'' shift
\begin{equation}
    \label{eqn:t_tilde}
    \tilde{t} = \int_0^T \frac{{\rm{d}}T^\prime}{ \tau\left(\sigma_{\rm{eff}}(T^\prime)/m\right)}.
\end{equation}
We denote $\tilde{t}$ as the adaptively scaled time and $t$ as the linear scaled time. Typically, $\tilde{t} \sim 1$ corresponds to the onset of core collapse when the core density increases exponentially. We define the adaptive collapse timescale $\tilde{\tau} (\sigma_{\rm{eff}}/m)$ as when
\begin{equation}
    \label{eqn:adaptive_collapse_timescale}
    1 = \int_0^{\tilde{\tau} (\sigma_{\rm{eff}}/m)} \frac{{\rm{d}}T^\prime}{ \tau\left(\sigma_{\rm{eff}}(T^\prime)/m\right)}.
\end{equation}

The effective cross section is the key quantity that encapsulates the physics of heat transfer in SIDM halos. It is typically expressed as a velocity averaging $\sigma_{\rm{eff}} = \left\langle \sigma_V (v_{\rm{rel}}) \, v_{\rm{rel}}^n \right\rangle / \left\langle v_{\rm{rel}}^n \right\rangle$, where the brackets signify the expected values assuming the isotropic Maxwell-Boltzmann (MB) velocity distribution. Following \cite{Yang+2022,Yang+23}, we utilize the heat conductivity-averaged cross section $\sigma_{\kappa} \equiv \langle \sigma v^5 \rangle / \langle v^5 \rangle $ with $n = 5$. The distribution of relative velocities in the kernel for $\sigma_{\rm{eff}}$ is therefore 
\begin{equation}
    \label{eqn:p_kappa}
    p_\kappa (v_{\rm{rel}}) \propto v_{\rm{rel}}^7 \, \exp{(-v_{\rm{rel}}^2/4 \sigma_{\rm{1D}}^2)},
\end{equation}
where $\sigma_{\rm{1D}}$ is the one-dimensional velocity dispersion of the core.  We will refer to the distribution in Equation \ref{eqn:p_kappa} as the conductivity-weighted MB distribution. The one-dimensional velocity dispersion is approximately given by $\sigma_{\rm{1D}} \approx 1.10 \, \sigma_{\rm{s}}$\footnote{Typically in the literature, the one-dimensional velocity dispersion of the core is approximated as $\sigma_{\rm{1D}} \approx 1.05 \, \sigma_{\rm{s}}$~\citep[e.g.][]{Yang+2022,Yang+23,Gilman+23}. This approximation remains valid early into the halo evolution; however, as the core continues to heat up in the core-collapse regime, taking $\sigma_{\rm{1D}} \approx 1.10 \, \sigma_{\rm{s}}$ becomes more appropriate.} for the calculation of the heat conductivity-averaged cross sections $\sigma_{\kappa}$ and the heat conductivity-weighted MB distributions $p_\kappa (v_{\rm{rel}})$. The linear scaled times are calculated using this approximation. Under this weighting, the most probable relative velocity is $v_{\rm{rel},\kappa}^{\rm{max}} = \sqrt{14} \, \sigma_{\rm{1D}}$, which we employ to match halo mass with relative velocity, as shown in Figure \ref{fig:cross_sections}. This relative velocity, along with the full width at half maximum (FWHM) of the heat conductivity-weighted MB distribution, can be used to characterize the relevant relative velocity range of a halo, especially as $\log M_{200}$ approximately depends linearly on $\log v_{\rm{rel},\kappa}^{\rm{max}}$. We adopt the halo mass-concentration relation in \cite{DiemerJoyce2019}.

Figure \ref{fig:cross_sections} shows the two SIDM models introduced in Section \ref{sec:particle_physics} alongside $\sigma_\kappa/m$. The FWHM of the conductivity-weighted MB distribution for the halos of mass $10^7, 10^8$, and $10^9\msun$ are also displayed as vertical shaded bars. Our simulation suite samples three main halo masses in this range, $M_{200} = 10^7, 10^8$, and $10^9\msun$, which have particle velocity distributions that span the most pronounced features of the RCS models. To illustrate this, in Figure \ref{fig:relative_velocity_distribution} we show the distribution
\begin{equation}
    \label{eqn:p_Gamma}
    p_\Gamma (v_{\rm{rel}}) \propto \Gamma (v_{\rm{rel}}) \, v_{\rm{rel}}^2 \, \exp{(-v_{\rm{rel}}^2/4 \sigma_{\rm{1D}}^2)},
\end{equation}
for each cross section model using the dashed curves. Here, $\Gamma (v_{\rm{rel}}) \propto \sigma_V (v_{\rm{rel}}) \, v_{\rm{rel}}$ represents the particle collision rate\footnote{Such distribution can also be employed to define a collision rate-averaged effective cross section, following $\sigma_\Gamma = \left\langle \sigma_V (v_{\rm{rel}}) \, \Gamma (v_{\rm{rel}}) \right\rangle / \left\langle \Gamma (v_{\rm{rel}}) 
\right\rangle$, which we studied in Paper\,I~\citep{Tran2024_1}}, and $p_\Gamma (v_{\rm{rel}})$ therefore represents the probability of particle collisions happening at the relative velocity range of $\left[ v_{\rm{rel}}, v_{\rm{rel}} + {\rm{d}}v_{\rm{rel}} \right]$. 

The distribution $p_\Gamma (v_{\rm{rel}})$ provides insight into which resonant features are the most influential for a halo's evolution. We show $p_\Gamma (v_{\rm{rel}})$ for each of the three halo masses and each RCS model in Figure \ref{fig:relative_velocity_distribution}. The cross sections are shown as shaded regions, the light dashed curve shows $p_{\Gamma}$ for the multi-peak RCS model, and the darker dashed curve shows $p_{\Gamma}$ for the single-peak RCS model. For the single-peak RCS model, the resonant peak remains highly influential across the three halo masses -- especially for the halo of mass $10^8\msun$, where most of the collisions are expected to happen near the resonance. For the multi-peak model, the collision rate-weighted MB distribution $p_{\Gamma}$ sweeps through different regions of the cross section profile, significantly changing the principle resonant features governing particle collisions.

\subsection{Numerical setup}
\label{subsec:numerical_setup}

\begin{table}
    \centering
    \addtolength{\tabcolsep}{0pt}
    \def\arraystretch{1.6}
    \begin{tabular}{c c c c c c c c}
        \hline
        $\log M_{200}$ & $N_{200}$ & $m_{\rm DM}$ & $r_{200}$ & $c$ & $r_{\rm{s}}$ & $\log \rho_{\rm{s}}$ & $\epsilon$ \\ [0.25ex] %
        [${\rm M}_\odot$] &  & [${\rm M}_\odot$] & [$\rm kpc$] &  & [$\rm kpc$] & [${\rm M}_\odot\,{\rm kpc}^{-3}$ ] & [$\rm pc$] \\ [1ex] %
        \hline\hline

        7 & $3 \!\times\! 10^7$ & 0.33 & 4.55 & 21.21 & 0.21 & 7.57 & 0.50 \\
        \hline
        
        7.5 & $3 \!\times\! 10^7$ & 1.05 & 6.68 & 19.81 & 0.34 & 7.50 & 0.71 \\
        \hline
        
        8 & $3 \!\times\! 10^7$ & 3.33 & 9.81 & 18.42 & 0.53 & 7.42 & 1.00 \\
        \hline
        
        8.5 & $3 \!\times\! 10^7$ & 10.5 & 14.4 & 17.05 & 0.84 & 7.33 & 1.41 \\ 
        \hline
        
        9 & $3 \!\times\! 10^7$ & 33.3 & 21.1 & 15.69 & 1.35 & 7.24 & 2.00 \\
        \hline
    \end{tabular}
    \caption{Simulations configuration. (1) $M_{200}$ is the virial mass of the halo. (2) $N_{200}$ is the number of DM particle within the virial radius (4) $r_{200}$. (3) $m_{\rm DM}$ is the mass of DM particles. (5) $c$ is the halo concentration parameter. (7) $\rho_{\rm{s}}$ and (6) $r_{\rm{s}}$ are the scale density and radius of the NFW profile. (8) $\epsilon$ is the (Plummer-equivalent) gravitational softening length of DM particles.}
    \label{tab:runs_config}    
\end{table}

\begin{table}
    \centering
    \addtolength{\tabcolsep}{7pt}
    \def\arraystretch{1.4}
    \begin{tabular}{c c c c c}
        \hline
        & \multicolumn{2}{c}{Single-peak} & \multicolumn{2}{c}{Multi-peak} \\
        $\log M_{200}$ & $\sigma_\kappa/m$ & $\tau_\kappa$ & $\sigma_\kappa/m$ & $\tau_\kappa$ \\ [0ex]
        [${\rm M}_\odot$] & [${\rm cm}^2\,{\rm g}^{-1}$] & [Gyr] & [${\rm cm}^2\,{\rm g}^{-1}$] & [Gyr] \\ [1ex]
        \hline\hline

        7 & 24.94 & 91.48 & 19.81 & 115.17 \\
        \hline
        
        7.5 & 43.54 & 43.39 & 19.76 & 95.6 \\ 
        \hline
        
        8 & 28.07 & 56.33 & 21.94 & 72.07 \\
        \hline
        
        8.5 & 14.62 & 91.73 & 18.60 & 72.1 \\ 
        \hline
        
        9 & 8.87 & 129.93 & 13.09 & 88.04 \\
        \hline
    \end{tabular}
    \caption{The heat conductivity-averaged cross section $\sigma_\kappa$ of the single- and multi-peak RCS models for each halo mass and the corresponding collapse timescale, calculated from Equation \ref{eqn:collapse_timescale} with $C=0.85$.}
    \label{tab:sigma_kappa}  
\end{table}

With the mass range identified, we perform idealized dark-matter-only N-body simulations of isolated halos using the multi-physics, massively parallel simulation code \textsc{Arepo} \cite{Springel2010,Weinberger2020}. For all simulations, gravity is computed using the Tree-Particle Mesh (Tree-PM) method with the softening lengths detailed in Table \ref{tab:runs_config}. DM self-interactions are modeled following the implementation in \cite{Vogelsberger2012}. All halos are initialized using stable configurations of halos in the CDM model, generated following the same procedure as detailed in~\citep{Tran2024_1}. The halos follow the classical NFW profile within the virial radius ($r_{200}$) and a modified exponential cut-off outside $r_{200}$~\citep{Springel1999}. The profiles take the analytical forms of
\begin{align}
    \label{eqn:nfw_inner_density}
    \rho_{r < r_{200}} (r) &= \frac{\rho_{\rm{s}}}{\left(r/r_{\rm s}\right) \left(1 + \left(r/r_{\rm s}\right)\right)^2}, \\
    \label{eqn:nfw_outer_density}
    \rho_{r > r_{200}} (r) &= \frac{\rho_{\rm s}}{c \left(1 + c\right)^2} \left(\frac{r}{r_{200}}\right)^{\epsilon_{\rm d}} \exp\left(-\frac{r-r_{200}}{r_{\rm d}}\right),
\end{align}
with $\rho_{\rm{s}}$ and $r_{\rm{s}}$ as the NFW scale density and radius. $c = r_{200} / r_{\rm{s}}$ is the halo concentration parameter determined using the $c \text{--} M$ relation. $r_{\rm d}$ is the decay scale, taken to be $r_{200}$ for simplicity. $\epsilon_{\rm d}$ is the exponential decay index chosen so that the continuity of the logarithmic slope of the density profile is preserved
\begin{equation}
    \label{eqn:nfw_decay_exponential}
    \epsilon_{\rm d} = \frac{r_{200}}{r_{\rm d}} - \frac{1+3c}{1+c}.
\end{equation}
As noted in Table \ref{tab:runs_config}, we resolve each halo with over 30 million particles. 

Our simulation suite include a total of 10 idealized N-body simulations with the halo masses of $10^7 M_{\odot}$, $10^{7.5} M_{\odot}$, $10^8 M_{\odot}$, $10^{8.5} M_{\odot}$, and $10^9 M_{\odot}$. For each halo mass, we evolve one halo in the single-peak resonant cross section and another in the multi-peak resonant cross section (RCS) model. The halos of mass $10^7 M_{\odot}$, $10^8 M_{\odot}$, and $10^9 M_{\odot}$ are evolved deep into the core-collapse regime $t \sim 1$, while we treat the halos of mass $10^{7.5}$ and $10^{8.5}\msun$ as supplements, only progressing the halos up to the end of the core-formation phase. All simulated halo configurations are detailed in Table \ref{tab:runs_config}. 

%These halo masses are chosen because their central velocity dispersions span the range of velocity scales where resonances appear in the scattering cross sections. 

In Section \ref{sec:results}, we will assess whether the halo density profile evolution is self-similar for each halo mass and each RCS models. Here, the self-similarity refers to a feature of SIDM models in which every halo follows a near-universal trajectory in central density vs. time, provided time is expressed in units of a characteristic self-interaction timescale (defined in the previous section). To quantify the degree to which resonant cross sections cause a deviation from self-similarity, for each halo mass, we also evolve two halos in the velocity-independent cross section (VICS) model, serving as benchmark simulations. The numerical values of the VICSs are taken as the heat conductivity-averaged cross section $\sigma_\kappa/m$ of the RCS models, the exact values of which are detailed in Table \ref{tab:sigma_kappa} along with the respective collapse timescales $\tau_\kappa$.

\subsection{Simulation analysis}
\label{subsec:analysis}

From the simulations, we obtain the halo density $\rho (r)$ and one-dimensional velocity dispersion $\sigma_{\rm{1D}} (r)$ profiles~\footnote{We compute the profiles by binning particles into 100 log-linearly spaced radial bins from $0.01\,r_{\rm{s}}$ to $3\,r_{200}$. We then merge neighboring bins with low particle counts to ensure a minimum count of $400$, corresponding to a minimum signal-noise ratio of $\sim 20$.}. From the initial NFW configuration, the central region of an SIDM halo will evolve into the isothermal state with a constant $\sigma_{\rm{1D}}$, and a smaller constant-density region (``core'') forms. Two approaches can be taken to determine this constant core density (and other related parameters). Here, we detail the methodology of the direct measurement approach, the results of which are shown in Section \ref{sec:results}, while Appendix~\ref{apd:core_evolution_fitted} presents the results based on density profile fitting~\citep{Tran2024_2}. To measure the constant core density, we start from the first radial bin $i$ with the cumulative particle count of $N_{\rm i} \geq 1000$ and calculate the cumulative average density $\bar\rho_i \pm \Delta{\bar\rho_i}$, following Poisson statistics. If $\lvert \bar\rho_i - \rho_{i+1} \rvert \, \geq 2 \left(\Delta{\bar\rho_i^2}+\Delta{\rho_{i+1}^2}\right)^{1/2}$, take $\bar \rho_i$ as the core density $\rho_{\rm{c}}$, else, consider the radial bin $i+1$ and repeat. For the core one-dimensional velocity dispersion $\sigma_{\rm{1D,c}}$, we repeat the same process, but using
\begin{equation}
    \label{eqn:core_velocity_dispersion}
    \bar \sigma_{{\rm{1D}},i} = \frac{\sum_{j<i} \sigma_{{\rm{1D}},j} / \Delta{\sigma_{{\rm{1D}},j}}^2}{\sum_{j<i} 1 / \Delta{\sigma_{{\rm{1D}},j}}^2}, \hspace{0.25cm} \Delta{\bar \sigma_{{\rm{1D}},i}} = \frac{1}{\sum_{j<i} 1 / \Delta{\sigma_{{\rm{1D}},j}}^2},
\end{equation}
as the cumulative average value. We calculate $\sigma_{{\rm{1D}},j}$ as the standard deviation of the one-dimensional velocities of particles within radial bin $j$. Assuming the one-dimensional velocity in each bin follows a normal distribution, the variance of the one-dimensional velocity will follow a chi-square distribution of order $3N$, where $N$ is the number of particles within the radial bin. In this case, the uncertainty of $\sigma_{{\rm{1D}},j}$, $\Delta\,\sigma_{{\rm{1D}},j}$, can be computed from this chi-square distribution. For the core one-dimensional velocity dispersion, we impose a larger limit for the cumulative particle count ($N_{\rm i} \geq 10000$) to find the first radial bin, while ignoring particles in the two innermost radial bins in order to reduce the impact of the highly fluctuating region deep inside the halo core.

In addition to the core density and core one-dimensional velocity dispersion, we also characterize the halo core with the core half-density radius $r_{\rho/2}$. As its name suggests, $r_{\rho/2}$ represents the radius at which the halo density drops to half the value at the core. Its uncertainty is calculated following a bootstrapping approach, that is, perturbing the density profile according to the measured statistic and recalculating $r_{\rho/2}$ for each of the samples. $\Delta r_{\rho/2}$ is then obtained as the standard deviation of the bootstrapped values. The core density and core half-density radius, along with their uncertainties at the epoch of maximum core size, are obtained using similar bootstrap procedures.

% In addition to the core density and velocity dispersion, we also parameterize the halo density profiles using the analytical profile proposed in \cite{Tran2024_2} with the form
% \begin{equation}
%     \label{eqn:TVS_density_profile}
%     \rho_{\rm{T24}} (r) = \rho_{\rm{c}} \left( \frac{\tanh{r/r_{\rm{c}}}}{r/r_{\rm{c}}} \right)^n \frac{1}{\left(1 + \left(r/r_{\rm{s}}^{\prime}\right)^2\right)^{\left(3-n\right)/2}}.
% \end{equation}
% Here, $r_{\rm c}$ is the characteristic core radius, and $r_{\rm{s}}^{\prime}$ is a scale radius similar to (but not to be confused with) the NFW scale radius $r_{\rm{s}}$. $n$ controls the steepness of the transition from the constant-density core ($\rho \sim \text{const}$) and the NFW tail ($\rho \sim r^{-3}$). As demonstrated in~\citep{Tran2024_2}, the profile has the advantage of closely approximating both the flat-core (i.e. $\rho \sim \text{const}$) and isothermal-core (i.e. $\sigma_{\rm{1D}} \sim \text{const}$) configurations in the inner regions, serving as an accurate parametric form for the halo density profile. We observe that, even when allowed to vary, the fitted values of $\rho_{\rm c}$ maintains consistency with the direct measurements described above. Nevertheless, we will measure $\rho_{\rm c}$ directly from the simulations, and use the parametric profile to examine the evolution of the core size $r_{\rm{c}}$ with time. 

\section{Results}
\label{sec:results}

\subsection{Evolution of halo structures}
\label{subsec:halo_evolution}

\begin{figure*}
    \centering
    \includegraphics[width=0.99 \textwidth]{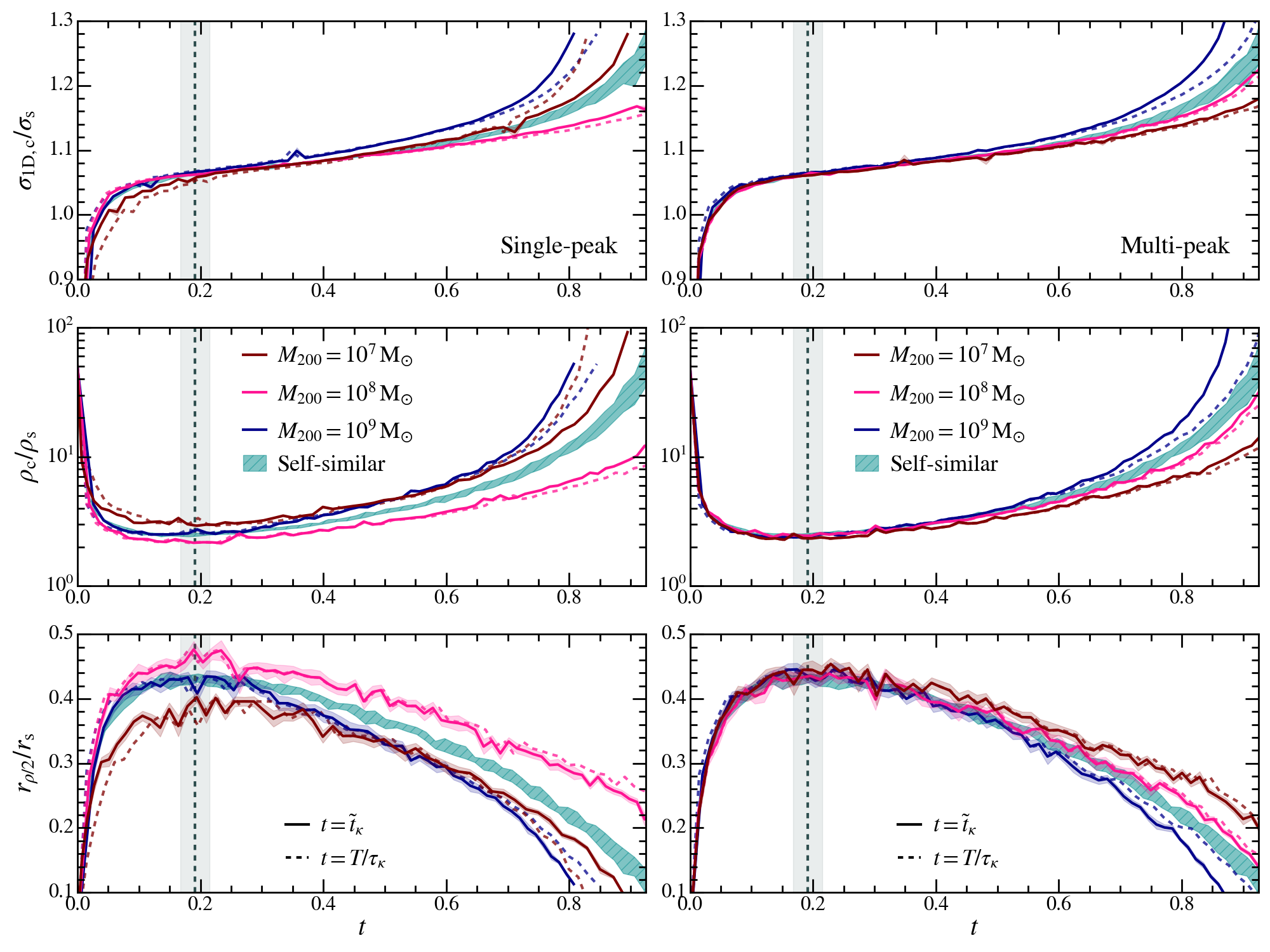}
    \caption{Evolution of the core one-dimensional velocity dispersion $\sigma_{\rm{1D,c}}$ (top), core density $\rho_{\rm{c}}$ (middle), and core half-density radius $r_{\rho/2}$ (bottom) of three halos in the single-peak (left) and multi-peak (right) RCS models in terms of the heat conductivity-averaged adaptive $\tilde{t}_\kappa$ (solid) and linear $t = T/\tau_\kappa$ (dashed) scaled times. The shaded regions surrounding the solid lines represent the measured uncertainties of the different parameters. The cyan-shaded region shows the $\pm 1\sigma$ variation in the universal self-similar, halo mass-independent core-collapse track in VICS models. The vertical dashed lines and the shaded gray regions indicate roughly when maximum core sizes are reached and the uncertainties. In terms of $\tilde{t}_\kappa$, the core-formation durations in different halos remain consistent with the universal evolution track. In this phase, the evolutions of the core one-dimensional velocity dispersion follow closely those in halos evolved under VICSs. In the core-collapse phase, halos evolved under the single-peak RCS model exhibit deviations of the order of $\sim 10 \text{--} 20\%$ from the universal self-similar core-collapse track, most clearly in the minimum core densities, maximum core radii, and the core-collapse times. On the other hand, halos evolved under the multi-peak RCS model follow the self-similar track closely, with deviations appearing only later in the core-collapse phase.}
    \label{fig:halo_evolutions}
\end{figure*}

\begin{figure}
    \centering
    \includegraphics[width= 0.49 \textwidth]{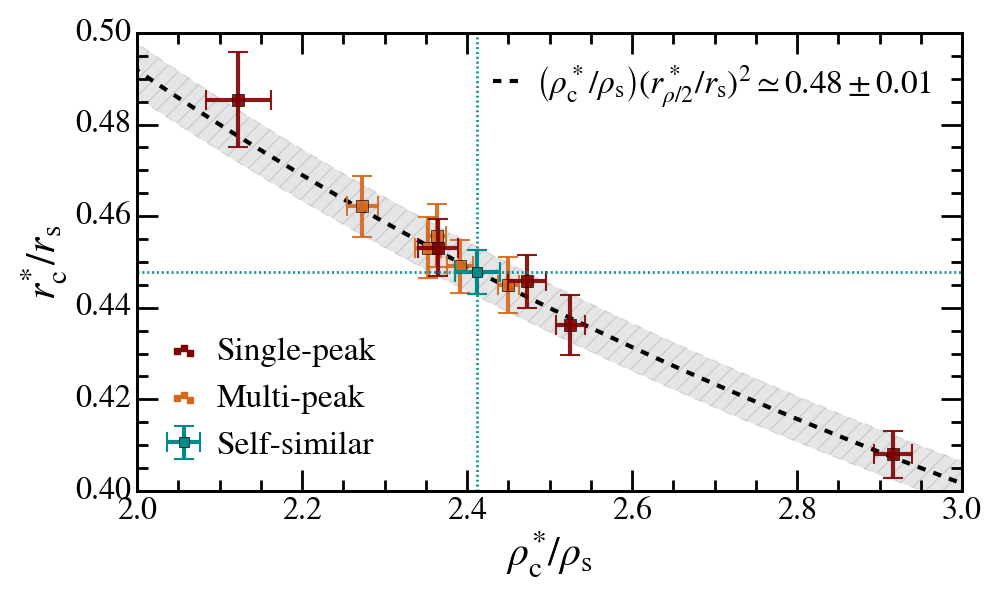}
    \caption{The dependency of the core half-density radius $r_{\rho/2}^{*}$ on core density $\rho_{\rm{c}}^{*}$ at the epoch of maximum core half-density radius in the single-peak (red) and multi-peak (orange) RCS models. The value achieved in a halo evolved under a VICS (cyan) is shown as a reference. As expected from core temperatures reaching similar (scaled) values across models, $r_{\rho/2}^{*}$ and $\rho_{\rm{c}}^{*}$ appear to be correlated following $\left(\rho_{\rm{c}}^{*}/\rho_{\rm{s}}\right) (r_{\rho/2}^{*}/r_{\rm{s}})^2 \sim \left(\sigma_{\rm{1D,c}}^{*}/\sigma_{\rm{s}}\right)^2 \simeq \text{const}$. The black line and gray-shaded region display such a relation and its corresponding uncertainty.}
    \label{fig:core_formation}
\end{figure}

\begin{figure}
    \centering
    \includegraphics[width= 0.49 \textwidth]{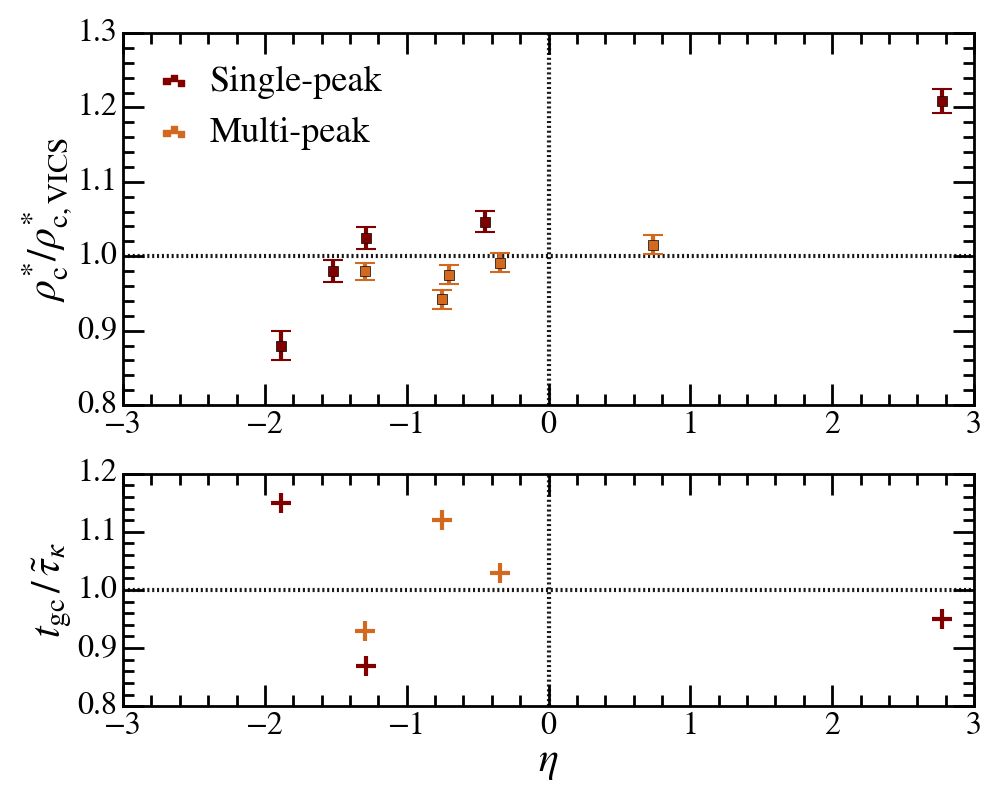}
    \caption{The minimum core density $\rho_{\rm{c}}^{*}$ (top) and core collapse time $t_{\rm{gc}}$ (bottom) (relative to their expected values in the VICS models) as a function of $\eta$ (Equation \ref{eqn:eta}). Results in the single- and multi-peak RCS models are shown in red and orange. Regarding $\rho_{\rm{c}}^{*}$, a positive correlation with $\eta$ is observed, suggesting that halos with greater variations in the effective cross section during evolution deviate more significantly from self-similarity. However, the core collapse time displays no clear correlation with $\eta$.}
    \label{fig:eta}
\end{figure}

\begin{table*}
    \centering
    \addtolength{\tabcolsep}{12pt}
    \def\arraystretch{1.6}
    \begin{tabular}{c c c c c c c}
        \hline
        & \multicolumn{3}{c}{Single-peak} & \multicolumn{3}{c}{Multi-peak} \\ [0.25ex]
        $M_{200} \, [\msun]$ & $10^7$ & $10^8$ & $10^9$ & $10^7$ & $10^8$ & $10^9$ \\ [1ex]
        \hline\hline

        $\eta$ & 2.78 & -1.89 & -1.29 & -0.75 & -0.35 & -1.29 \\
        \hline

        $\rho_{\rm{c}}^{*}/\rho_{\rm{c,VICS}}^{*}$ & $1.21 \pm 0.01$ & $0.88 \pm 0.02$ & $1.02 \pm 0.01$ & $0.94 \pm 0.01$ & $0.99 \pm 0.01$ & $0.98 \pm 0.01$ \\
        \hline

        $t_{\rm{gc}} / \tilde{\tau}_\kappa$ & 0.95 & 1.15 & 0.87 & 1.12 & 1.03 & 0.93 \\
        
        $t_{\rm{gc}} / \tau_\kappa$ & 0.90 & 1.25 & 0.90 & 1.15 & 1.05 & 0.97 \\
        \hline
        
    \end{tabular}
    \caption{Top: The values of $\eta$ (Equation \ref{eqn:eta}) for the halos of mass $10^7\msun$, $10^8\msun$, and $10^9\msun$ in the single- and multi-peak RCS models. Middle: Ratios between the minimum core density of halos evolved under RCSs and the expected value in VICS cases. Bottom: The approximated ratios between the observed collapse times of halos evolved under RCS models and the predicted heat conductivity-averaged adaptive $\tilde{\tau}_\kappa$ (Equation \ref{eqn:adaptive_collapse_timescale}) and linear $\tau_\kappa$ (Equation \ref{eqn:collapse_timescale}) collapse timescales.}
    \label{tab:eta}    
\end{table*}

In Figure \ref{fig:halo_evolutions}, we show the time evolution of the core one-dimensional velocity dispersion $\sigma_{\rm{1D,c}}$, core density $\rho_{\rm{c}}$, and core half-density radius $r_{\rho/2}$ for three halos in the single-peak and multi-peak RCS models. They are compared to the universal self-similar core-collapse solution found in runs with VICSs. As the simulation results of these runs exhibit slight deviations due to numerical effects~\citep{Mace+24,Palubski2024}, we present the self-similar core-collapse solution as the $\pm 1\sigma$ regions aggregated from the results of such runs. We normalize the evolutions with the initial NFW profile parameters $\sigma_{\rm{s}}$, $\rho_{\rm{s}}$, and $r_{\rm{s}}$, while characterizing the halo evolutions with timescales calculated using the heat conductivity-averaged effective cross sections $\sigma_\kappa$\footnote{In Paper\,I~\citep{Tran2024_1}, this was found to give a good description of halo evolution in the core-collapsed phase. For the core-formation phase, we found that an alternative collision rate-averaged effective cross section provided a better description, but this does not generalize to the halos with varying masses in this paper.}. When the halo evolutions are expressed in terms of the corresponding adaptively scaled time $\tilde{t}_\kappa$, we find that the core-formation times are invariant of halo mass and consistent between VICS and RCS models. This happens around $\tilde{t}_\kappa \sim 0.2$ when the core is thermalized and the core size $r_{\rm c}$ reaches the maximum. The evolutions of $\sigma_{\rm{1D,c}}$ appear self-similar across all halos up to $\tilde{t}_\kappa \sim 0.4$, far beyond the epoch of core formation. The halo core typically enters the core-collapse phase after the epoch. 

In Figure~\ref{fig:core_formation}, we show the dependency of the core half-density radius $r_{\rho/2}^*$ on the core density $\rho_{\rm{c}}^*$ at the epoch of maximum core half-density radius for halos in both single- and multi-peak RCS models, including the halos of mass $10^{7.5}$ and $10^{8.5}\msun$. We observe a correlation between these parameters, i.e. $\left(\rho_{\rm{c}}^{*}/\rho_{\rm{s}}\right) (r_{\rho/2}^{*}/r_{\rm{s}})^2 \simeq \text{const}$. This is consistent with the fact that the one-dimensional velocity dispersion of the core at the epoch reach a constant value of $\sigma_{\rm{1D,c}}^{*} \sim 1.05 \, \sigma_{\rm{s}}$ in both VICS and RCS models, although $r_{\rho/2}^*$ and $\rho_{\rm{c}}^*$ can vary. The relation $\left(\rho_{\rm{c}}^*/\rho_{\rm{s}}\right) (r_{\rho/2}^*/r_{\rm{s}})^2 \simeq 0.48 \pm 0.01$, calculated from the self-similar values ($\rho_{\rm{c,VICS}}^*$ and $r_{\rho/2\rm{,VICS}}^*$), are displayed in Figure \ref{fig:core_formation}. The corresponding uncertainty is propagated from the uncertainties of $\rho_{\rm{c,VICS}}^*$ and $r_{\rho/2\rm{,VICS}}^*$.

For the single-peak model, we find that the minimum core density and maximum core half-density radius depend strongly on the halo mass or, equivalently, on the alignment of the halo velocity structure to the resonant peak. The values achieved in the halo of mass $10^9\msun$ remain consistent with those in halos evolved under VICSs. As shown by Figure \ref{fig:cross_sections}, this occurs because the halo central velocity dispersion is significantly offset from the position of the resonance. However, we observe a smaller but denser core in the halo of mass $10^7\msun$, while the halo of mass $10^8\msun$ develops a larger but more diffuse core, despite these halos having comparable heat conductivity-averaged effective cross sections. These deviations are of the order of $\sim 10 \text{--} 20\%$, compared to values achieved in the VICS cases. 

We can make some headway towards understanding the deviations from self-similar evolution by examining the change in the effective cross section with respect to a shift in the core one-dimensional velocity dispersion, characterized by
\begin{equation}
    \label{eqn:eta}
    \eta = \frac{\partial \ln{\sigma_\kappa}}{\partial \ln{\sigma_{\rm{1D,c}}}}.
\end{equation}
For VICS models, $\eta=0$; however, for RCS models, we expect significant changes in $\eta$ as the velocity distribution of particles inside a halo sweeps across a resonance. A positive value of $\eta$ indicates an increase in effective cross section as the halo core heats up. Table \ref{tab:eta} summarize the values of $\eta$ for the main halo masses in both single- and multi-peak RCS models. We observe the values of $\eta \simeq 2.78$ and $\eta \simeq -1.89$ for the halos of mass $10^7\msun$ and $10^8\msun$, respectively, while the halo of mass $10^9\msun$ exhibits the value of $\eta \simeq -1.29$. These values appear to be correlated to the values of the minimum core densities achieved in halos, as illustrated in Figure \ref{fig:eta}, where the ratios between values achieved in RCS runs and those in VICS runs are displayed.

Regarding the core-collapse phase, we observe shifts of approximately $\sim 10\text{--}15\%$ in the core-collapse tracks and the onset of the gravothermal catastrophe in halos evolved under the single-peak RCS model, compared to those in runs with VICSs. These results are consistent with and expand on the results presented by ~\citep{Tran2024_1}. To quantify when core collapse occurs relative a VICS model, we compute the gravothermal core collapse times of halos in our simulations $t_{\rm{gc}}$, with the predicted collapse time from the conductivity-averaged adaptive $\tilde{\tau}_\kappa = \tilde{\tau}(\sigma_\kappa/m)$ (Equation \ref{eqn:collapse_timescale}) and linear $\tau_\kappa = \tau(\sigma_\kappa/m)$ (Equation \ref{eqn:adaptive_collapse_timescale}) collapse timescales. Here, we take $t_{\rm{gc}}$ as the approximated physical time at which $\rho_{\rm{c}} \rightarrow \infty$ (and $r_{\rho/2} \rightarrow 0$). The results are summarized in Table \ref{tab:eta}, along with the other quantifiable deviations from the self-similar track of halos under RCSs. We see that in the single-peak RCS model, the halo of mass $10^8\msun$ suffers a deceleration in the collapse time, while the core-collapse processes in the halos of mass $10^7$ and $10^9\msun$ are accelerated. There appear to be no correlation between the acceleration of core collapse and $\eta$, contrary to what is observed in the case of the minimum core density.

On the other hand, in the multi-peak RCS model, the time evolutions of the three halos are almost identical to those in the VICS runs. Deviations from the self-similar tracks occur, but at a much later stage compared to in the single-peak model. These deviations are also of the order of $\sim 5\text{--}10\%$ in the collapse times, similar to what was observed in the single-peak RCS runs. One important difference between the multi-peak and single-peak models is the sensitivity of the effective cross section on the velocity distribution of DM particles as hinted by Figure \ref{fig:cross_sections}. In the multi-peak model, multiple resonant features with either low prominence or narrow width, compared to the typically halo relative velocity distributions, in close proximity suppress fluctuations and cause the effective cross section to depend weakly on the halo mass. This is most clearly observed in the halos of mass neighboring $10^8\msun$, where $\lvert \eta \rvert \simeq 0.35$. Other halos in the multi-peak RCS model typically exhibit the value of $\lvert \eta \rvert \lesssim 1$, compared to $\lvert \eta \rvert \sim 2$ for the more aligned halos in the single-peak model. At such values, the halos closely resemble those evolved under VICSs, and it is not surprising that the multi-peak RCS run results mimic those of the VICS runs.

\subsection{Diversity of the rotation curves from SIDM resonance}
\label{subsec:diversity}

\begin{figure*}
    \centering
    \includegraphics[width=0.99 \textwidth]{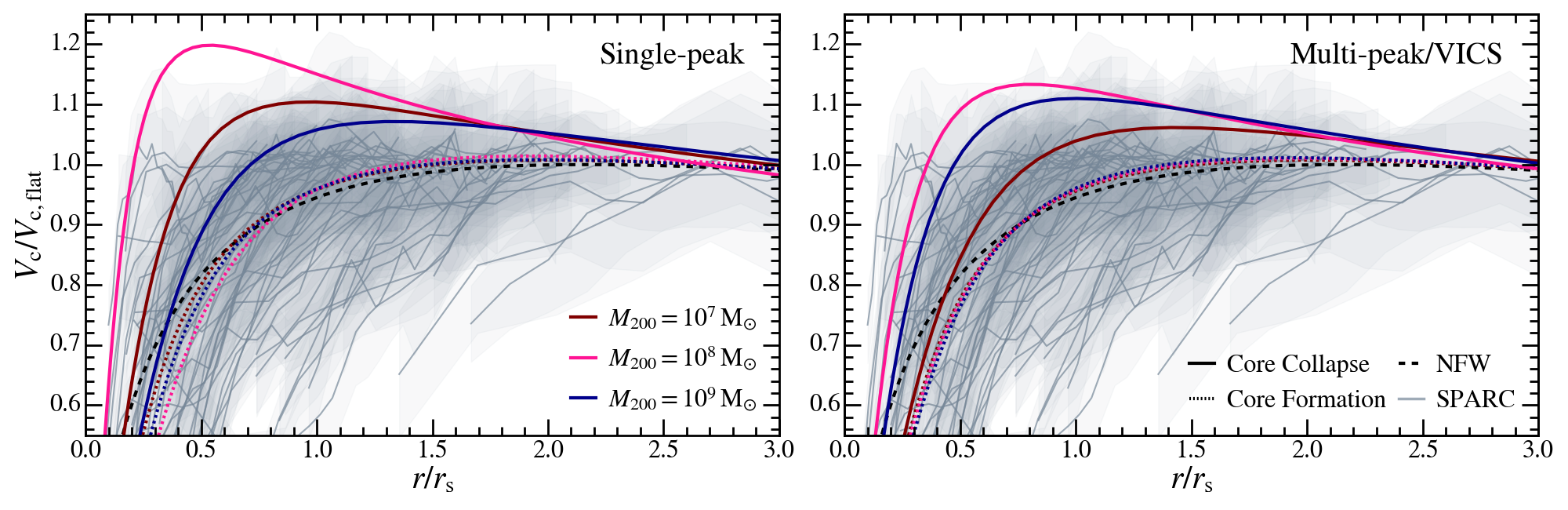}
    \caption{Rotation curves of simulated halos evolved under the single-peak (left) and multi-peak (right) RCS models in the core-formation $T \approx 3.6\Gyr$ (dotted) and core-collapse $T \approx 62.5\Gyr$ (solid) phase. The rotation curves of halos evolved in the multi-peak RCS model mimic the behavior expected in those under VICSs. The NFW rotation curve (dashed) is shown as a reference. Rotation velocities $V_{c}$ are normalized using the maximum value from the initial NFW profile, as substitute for the asymptotic flat rotation velocities $V_{c,\rm{flat}}$, while radii $r$ are normalized using the NFW scale radius $r_{\rm{s}}$. For comparison, we show the observed 52 highest-quality HI rotation curves from the SPARC project~\citep{Lelli2016_2} (gray) with uncertainties (shaded regions) propagated from the measurements of $V_{\rm{c}}$ and $V_{c,\rm{flat}}$. $V_{c,\rm{flat}}$ of these rotation curves are obtained from \cite{Lelli2016_1}, and $r_{\rm s}$ are estimated as described in Section \ref{subsec:diversity}. The rotation curves show a substantial diversity due to the evolution of the core-collapse process, with additional contributions from the deviations generated by the resonance of DM self-interaction. Most of the observed rotation curves fall within this range, except for several shallower ones, which may result from the unaccounted stellar feedback and environmental effects or imperfections in our normalizing approach.}
    \label{fig:rotation_curves}
\end{figure*}

One challenge confronted by the standard CDM paradigm is the observed diversity of dwarf galaxy rotation curves~\citep[see recent reviews in][]{Sales2022,Primack2024}. While a population of dwarf galaxies exhibits kpc-sized DM cores, others display cuspy central density distributions. The implication of SIDM for this problem has been investigated in literature~\citep[e.g.][]{Kamada2017,Creasey2017,Omid2020,Oman2015,Kaplinghat2019,Jiang2023,Shen2024} but focusing primarily on the impact of halo concentrations and gravitational potential of baryons on SIDM density profile. However, in SIDM models with RCS, an additional route to diversity emerges as small differences in halo velocity dispersion can yield large variations in the effective self-interaction cross section, thereby producing cored and core-collapsed halos at similar mass scales. 

In Figure \ref{fig:rotation_curves}, we compare the rotation curves ($V_{\rm c}\equiv \sqrt{G\,M_{\rm enc}(<r)/r}$, where $M_{\rm enc}$ is the enclosed mass) of simulated DM halos in different evolution phases to observations. We pick a snapshot when core-formation is in process for all halos ($T \approx 3.6\Gyr$) and another snapshot when the halos are deep into the core-collapse phase ($T \approx 62.5\Gyr$) to measure the rotation curves. The observed HI rotation curves are taken from the Spitzer Photometry and Accurate Rotation Curves (SPARC) project~\citep{Lelli2016_2}. We adopt the $V_{\rm c}$ contributed by DM from SPARC's dynamic modeling and the corresponding uncertainties, i.e., after subtracting the gas and stellar contribution from their mass modeling. Our simulated halos have significantly lower masses compared to the typical halo mass of these observed field dwarf galaxies ($\sim 10^{10}$--$10^{11}\msun$). Therefore, for simulated halos, we normalize the rotation velocities with the maximum rotation velocities of the initial NFW profiles (as proxies to the asymptotically flat rotation velocities $V_{\rm{c,flat}}$) and the radii with the NFW scale radius $r_{\rm{s}}$. Meanwhile, for the observed dwarfs, the rotation curves are normalized by the asymptotically flat rotation velocities $V_{\rm{c,flat}} \pm \Delta V_{\rm{c,flat}}$, calculated following~\citep{Lelli2016_1}. The radii are normalized by an estimated $r_{\rm{s}}$ assuming the mass-concentration relation~\citep{DiemerJoyce2019}. We include only 52 of the 175 observed galaxies in our analysis, leaving out those with increasing rotation curves (i.e., no asymptotically flat rotation velocity detected) and those with high uncertainties in the measurements. We note that the normalizations used here are approximated, and we aim mainly for a proof of concept here.%our goal is a qualitative comparison designed primarily to illustrate the key points, namely, that resonances can contribute the rotation curve diversity.
%We note that different $c \textup{--} M$ relation models produce different values of $r_{\rm{s}}$, resulting in different normalization of radii. Finding such an appropriate normalization of the rotation curves can provide insight into the proper relation between the concentration parameter and halo mass.
%The uncertainties in the scale radius $\sigma_{r_{\rm{s}}}$ are propagated from the uncertainties of the asymptomatically flat rotation velocity $\sigma_{V_{\rm{c,flat}}}$, with an additional factor of $1.2$ to account for the fact that the halos may have undergone core formation or core collapse. 

Diversity of rotation curves arises from SIDM halos being in different evolutionary stages. Halos with shorter collapse timescale evolve deeper in the core-collapse phase and obtain denser cores with cuspier profiles, while halos with longer collapse timescale are still in the core-formation phase and feature shallower rotation curves. We observe that for halos in the multi-peak RCS model, the difference in collapse timescale resulted mostly from the differences in the initial halo density $\rho_{\rm{s}}$ and velocity dispersion $\sigma_{\rm{s}}$, analogous to the behavior expected in halos evolved under VICSs. For halos in the single-peak RCS mode, variations in the effective cross section due to strong resonance further amplify the diversity of rotation curves. Halos with the relative velocity distribution (conductivity-weighted) more aligned with the resonant peak have a higher effective cross section and collapse faster. Meanwhile, halos that are misaligned with the resonant peak, resulting in a lower effective cross section and higher collapse timescale, remain in earlier stages. The magnitude of diversity from all these factors is most prominent in the core-collapse phase and is comparable to the observed diversity of dwarf galaxy HI rotation curves. However, a number of observed dwarfs have shallower rotation curves than those of the simulated SIDM halos, even in the core-formation stage. These rotation curves, nevertheless, could be the results of other physical processes, such as stellar feedback and environmental effects, which drive further diversity.

%the processes we have identified should also occur with resonant SIDM models. 

%Additional diversity could come from different concentrations of halos at the same mass scale, which can also change the effective SIDM cross section and alignment to the resonant features. 

% However, dwarf galaxies with particularly shallower HI rotation curves remain unexplained by the DM self-interacting evolution of halos. These rotation curves, nevertheless, are consistent with those exhibited by halos with isothermal cores, and high representativeness fits using the isothermal core profile $\rho_{\rm{T24}}$ (Equation \ref{eqn:TVS_density_profile}) with $n=2.5$ and constrained maximum rotation velocity of $V_{\rm{c,max}}=V_{\rm{c,flat}}$ can be achieved. This could hint at the necessity of remodeling the  $c \textup{--} M$ relation as mentioned above, in order to obtain more proper values of $r_{\rm{s}}$ for normalizing the rotation curves. An additional conclusion regarding the constraint of DM self-interactions can also be reached using the optimized $r_{\rm{c}}$ and $V_{\rm{c,max}}$ from the aforementioned fits. However, the implementation of such analyses would be complicated and require details consideration of different factors, such as the $c \textup{--} M$ relation model and the evolution tracks of DM halo under self-interactions.

\section{Discussion and conclusions}
\label{sec:conclusions}

In this work, we have explored the implications of SIDM models with resonant scattering cross sections on the evolution and structure of DM halos. We identify the most relevant halo mass range to the resonance features, using particle collisional velocity distributions that govern the heat conductivity and evolution of halos. We conduct a suite of high-resolution N-body simulations of halos in the mass range of $10^{7}\textup{--}10^{9}\msun$ in both single-peak and multi-peak RCS models. The results of these simulations are compared to the evolution of halos evolved under VICSs, as well as to observed rotation curves of field dwarf galaxies. We summarize our findings as follows:
\begin{itemize}
    \item When expressed in terms of the adaptive scaled time $\tilde{t}_\kappa$ (Equation \ref{eqn:t_tilde}), calculated using the heat conductivity-averaged effective cross section $\sigma_\kappa \equiv \langle \sigma v^5 \rangle / \langle v^5 \rangle$, the core-formation times of halos are invariant of halo mass and cross section models. For the core one-dimensional velocity dispersion $\sigma_{\rm{1D,c}}$, we observe self-similar evolutions consistent with the universal track in VICS runs, up to $\tilde{t}_\kappa \sim 0.4$, across halos in both RCS models. The core density and radius of halos at core formation also appear to correlate in a manner that preserves such velocity dispersions.
    
    \item For the single-peak RCS model, we see deviations of order of $\sim 10 \text{--} 20\%$ in the minimum core density and maximum core half-density radius for the halos of mass $10^7\msun$ and $10^8\msun$, the mass scales aligned with the prominent resonant feature of the single-peak cross section. On the other hand, in the multi-peak RCS model, the time-evolution more closely resembles the VICS evolution. The core-collapse times deviate at the order of $\sim 5 \text{--} 10\%$ from the VICS self-similar evolution for the multi-peak model, compared to $\sim 10 \text{--} 15\%$ in the single-peak model. 
    
    \item SIDM models with resonant features can produce a wide range of rotation curves in halos evolved for the same amount of time, depending on the effective cross section. In the single-peak RCS model, halos that are more aligned with the resonance develop denser cores and steeper rotation curves due to the higher effective cross section and shorter core-collapse timescale. At the same time, those misaligned with a resonance remain in earlier evolution stages with shallower rotation curves. The diversity resulting from the differences in the effective cross sections are generally larger than that in the multi-peak RCS model, which resemble the VICS cases.
\end{itemize}

Our results demonstrate that SIDM models with resonances can drive different DM halo evolution patterns compared to VICS cases. This is the same conclusion reached by \cite{Tran2024_1}, who analyzed the evolution of a $10^8 M_{\odot}$ halo evolving with the single-peak model. We have expanded on this previous work by considering a different RCS model with multiple peaks, and several halos spanning two decades in halo mass evolving with each cross section. However, we note that the deviations are only at the level of $\sim 20\%$ in the core-collapse tracks. This suggests that modeling halo evolution using a universal or self-similar model for the halo evolution \cite[e.g.][]{Yang2024} could remain accurate enough for many practical applications.

We can understand the deviation from self-similarity in the RCS models as follows: as the cores of our simulated halos heat up from self-interactions, the distribution of particle velocities move across resonant peaks in the cross section; the cross section strength seen by particles therefore changes with time, and also with the radius inside the halo. In Section \ref{sec:results}, we quantified this physical picture in terms of a parameter $\eta$, which captures the change of cross section strength with the one-dimensional velocity dispersion. The halos with larger $\eta$ tend to have more significant deviations from self-similar evolution. 

Halo configurations resulting from RCS models exhibit more diversity in their rotation curves relative to VICS models. It remains to be seen whether this increase in diversity persists when other relevant effects, such as tidal stripping, evaporation, and stellar feedback, also operate on a halo density profile. Investigating these possibilities would require cosmological simulations with hydrodynamics. Future work, incorporating extended mass ranges, variations in halo concentrations, and baryonic physics will further refine our understanding of SIDM and its role in shaping the universe's structure on small scales.

\begin{acknowledgments}
We thank the anonymous referee for useful comments and suggestions. The simulations in this paper were conducted on the Engaging cluster at Massachusetts Institute of Technology (MIT) operated by the MIT Office of Research Computing and Data. DG acknowledges support for this work provided by the Brinson Foundation through a Brinson Prize Fellowship grant. XS acknowledges the support from the National Aeronautics and Space Administration (NASA) theory grant JWST-AR-04814.
\end{acknowledgments}

% \nocite{*}

\bibliography{main}% Produces the bibliography via BibTeX.

\appendix

\section{Evolutions of the halo cores in terms of fitted parameters}
\label{apd:core_evolution_fitted}

\begin{figure*}
    \centering
    \includegraphics[width=0.99 \textwidth]{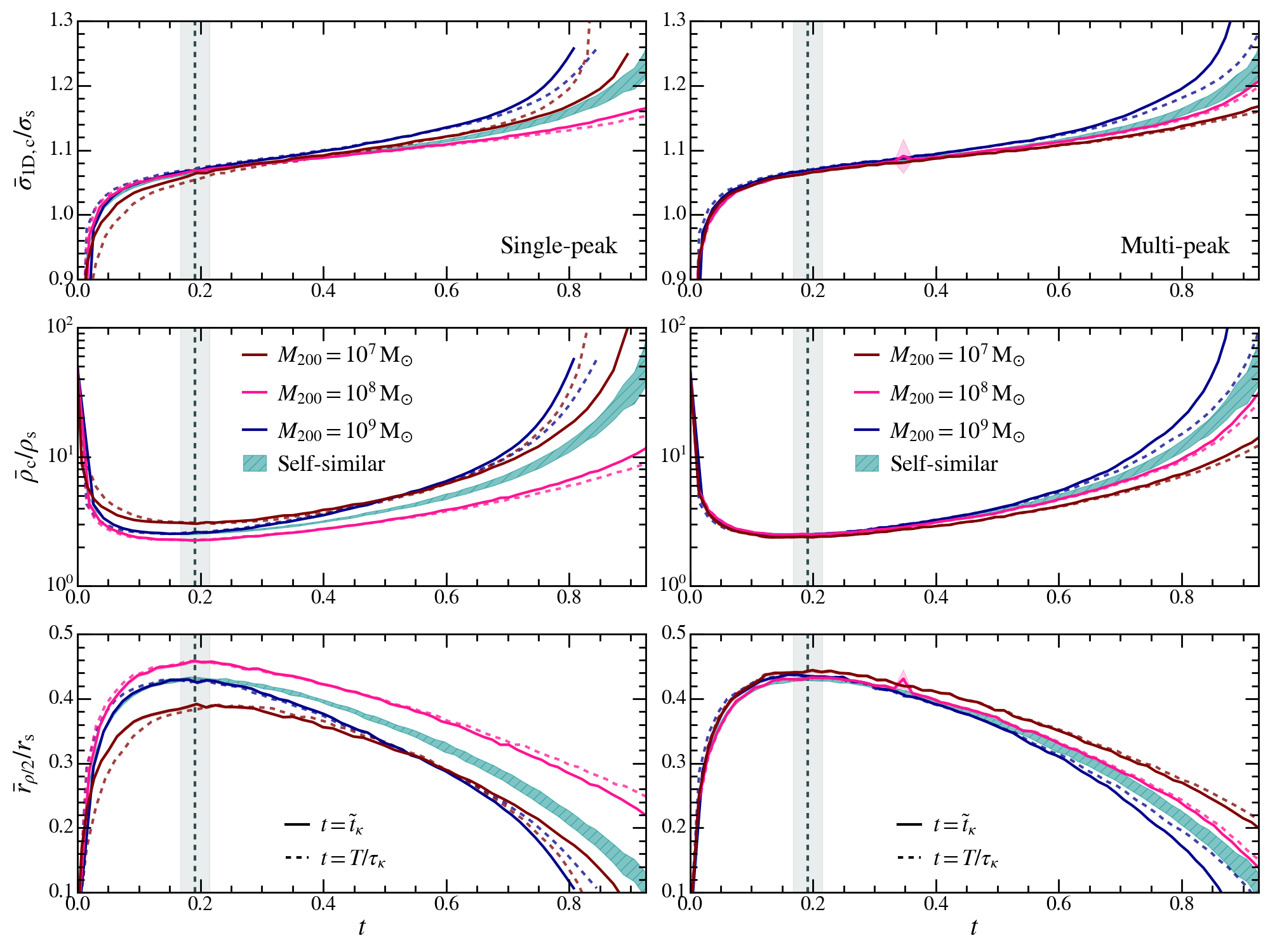}
    \caption{Similar to Figure \ref{fig:halo_evolutions}, but showing the fitted core one-dimensional velocity dispersion $\bar{\sigma}{\rm{1D,c}}$ (top), core density $\bar{\rho}{\rm{c}}$ (middle), and core half-density radius $\bar{r}_{\rho/2}$ (bottom) instead of direct measurements. Compared to Figure \ref{fig:halo_evolutions}, the parameter evolutions remain consistent, albeit appearing smoother, with reduced uncertainties, even in the case of the half-density radius. Overall, conclusions on halo evolutions are in agreement with those of the previous figure.}
    \label{fig:halo_evolutions_fitted}
\end{figure*}

\begin{figure}
    \centering
    \includegraphics[width= 0.49 \textwidth]{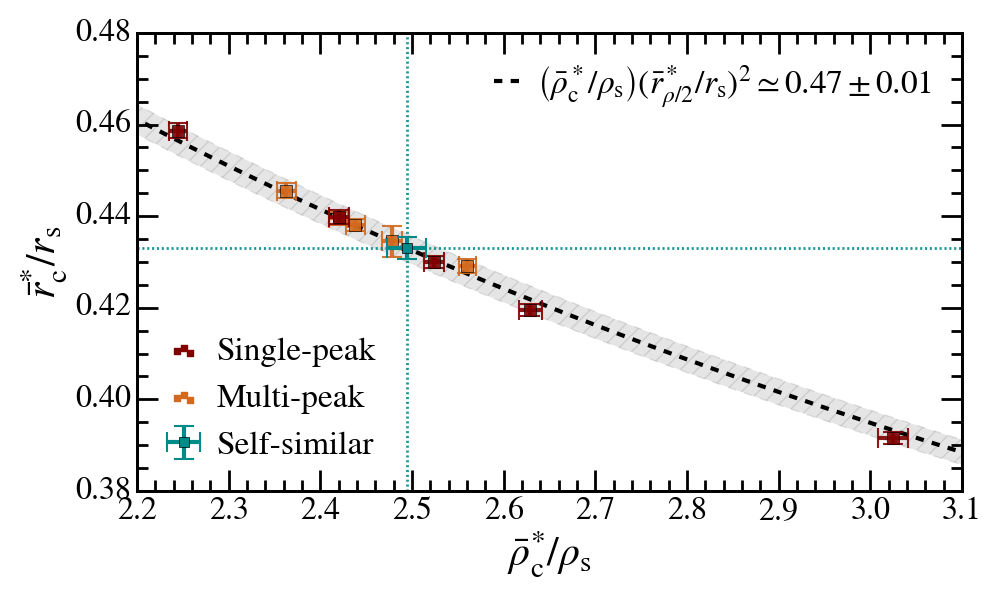}
    \caption{Similar to Figure \ref{fig:core_formation}, but showing the dependency of the fitted core half-density radius $\bar{r}_{\rho/2}^{*}$ on the fitted core density $\bar{\rho}_{\rm{c}}^{*}$ at the epoch of maximum core half-density radius. Here, a much higher agreement with the $\left(\rho_{\rm{c}}^{*}/\rho_{\rm{s}}\right) \left(r_{\rho/2}^{*}/r_{\rm{s}}\right)^2 \simeq \text{const}$ relation are reached, compared to in the previous figure.}
    \label{fig:core_formation_fitted}
\end{figure}

In addition to the algorithms detailed in Section \ref{subsec:analysis}, we measure the core density and half-density radius using the analytical profile proposed in \cite{Tran2024_2} with the form
\begin{equation}
    \label{eqn:TVS_density_profile}
    \rho_{\rm{T24}} (r) = \rho_{\rm{c}} \left( \frac{\tanh{r/r_{\rm{c}}}}{r/r_{\rm{c}}} \right)^n \frac{1}{\left(1 + \left(r/r_{\rm{s}}^{\prime}\right)^2\right)^{\left(3-n\right)/2}}.
\end{equation}
Here, $r_{\rm c}$ is the characteristic core radius, and $r_{\rm{s}}^{\prime}$ is a scale radius similar to (but not to be confused with) the NFW scale radius $r_{\rm{s}}$. $n$ controls the steepness of the transition from the constant-density core ($\rho \sim \text{const}$) and the NFW tail ($\rho \sim r^{-3}$). As demonstrated in~\citep{Tran2024_2}, the profile has the advantage of closely approximating both the flat-core (i.e. $\rho \sim \text{const}$) and isothermal-core (i.e. $\sigma_{\rm{1D}} \sim \text{const}$) configurations in the inner regions, serving as an accurate parametric form for the halo density profile. The uncertainties of the fitted parameters are calculated following a bootstrapping approach similar to that detailed in Section \ref{subsec:analysis}, i.e., perturbing the density profile according to the measured uncertainties, repeating the fitting process for each sample, and calculating the standard deviations of the resulting parameters. Onward, we utilize bars to differentiate fitted parameters from those measured following the algorithms in Section \ref{subsec:analysis}, e.g., $\bar{\rho}_{\rm{c}}$ compared to $\rho_{\rm{c}}$, or $\bar{\sigma}_{\rm{1D,c}}$ compared to $\sigma_{\rm{1D,c}}$.

The core half-density radius $\bar{r}_{\rho/2}$ can be calculated from the core radius $r_{\rm{c}}$ and the transition index $n$ following $r_{\rho/2} = r_{\rm{c}} f^{-1}(0.5^{1/n})$, with $f(x) = \tanh{x}/x$. After the core-formation stage, $n$ typically take the value of $2.5$, resulting in $f^{-1}(0.5^{1/n}) \simeq 1$. Meanwhile, the core one-dimensional velocity dispersion $\bar{\sigma}_{\rm{1D,c}}$ can be calculated as the average velocity dispersion within $r_{\rm{c}}$, following~\citep{Tran2024_2}, which shows values of $\bar{\sigma}_{\rm{1D,c}}$ serving as good representations for the isothermal central region, up to $T \sim 0.8 \tau$. Generally, the values of the fitted parameters and those measured following Section~\ref{subsec:analysis} remain consistent. Nevertheless, as shown in Figure~\ref{fig:halo_evolutions_fitted}, the halo evolution described in terms of the fitted parameters appears smoother, with trends appearing clearer. Overall, similar observations to Section~\ref{subsec:halo_evolution} can be made, especially by Figure~\ref{fig:core_formation_fitted}, where the $\left(\rho_{\rm{c}}^{}/\rho_{\rm{s}}\right) (r_{\rho/2}^{}/r_{\rm{s}})^2 \sim \left(\sigma_{\rm{1D,c}}^{*}/\sigma_{\rm{s}}\right)^2 \simeq \text{const}$ relation is even more strongly supported than by Figure~\ref{fig:core_formation}.

\end{document}